\documentclass[preprint,12pt]{aastex}
\usepackage{graphicx}
\usepackage{color,calc}
\usepackage{amsmath}
\usepackage{natbib}
\newcommand{\tbfrac}[2]{\genfrac{}{}{0pt}{0}{#1\strut}{#2\strut}}
\shorttitle{Disruption of Clusters near the Galactic Center}
\shortauthors{M.\ A.\ G\"urkan, F.\ A.\ Rasio}
\title{The Disruption of Stellar Clusters Containing Massive Black Holes near
the Galactic Center}
\author{M.\ Atakan G\"urkan and
Frederic A.\ Rasio}
\affil{Department of Physics and Astronomy, Northwestern University,
Evanston, IL 60208\\
{\tt ato, rasio@northwestern.edu}}
\begin{document}
\begin{abstract}
We present results from dynamical Monte Carlo simulations of dense star
clusters near the Galactic center. These clusters sink toward the center
of the Galaxy by dynamical friction. During their inspiral, they may
undergo core collapse and form an intermediate-mass black hole through runaway
collisions. Such a cluster can then reach within a parsec of the Galactic
center before it completely disrupts, releasing many young stars in this 
region. This scenario provides a natural explanation
for the presence of the young stars observed near the Galactic center. Here 
we
determine the initial conditions for this scenario to work and we derive
the mass distribution of cluster stars as a function of distance from the
Galactic center. For reasonable
initial conditions, we find that clusters massive enough for rapid inspiral would
include more massive stars ($m_\star \gtrsim 30\,M_\odot$) than currently
observed in the inspiral region. We point out several possible
explanations for this apparent discrepancy.  
\end{abstract}


\section{Introduction}
Multi-wavelength observations of the Galactic center (GC) have revealed
a supermassive black hole (SBH) and a stellar cusp surrounding it. 
The mass of the SBH, 
$M_{\rm SBH}\simeq 4\times10^6\,M_\sun$, is one of the most reliable
estimates for massive black holes at the centers of galaxies. The stellar population within 1\,pc of the SBH contains a variety
of young and massive stars. Some of them are only about 20\,Myr old and get
as close as a few light days to the SBH; while from 0.1 to 0.4\,pc 
even younger stars are found with ages 3 to 7\,Myr.
The presence of these young stars in the immediate vicinity of the SBH
poses a problem known as the {\em youth paradox\/}. Their {\em in situ\/}
formation is problematic since the SBH has a strong tidal influence in
this region. However, the
time required for the migration of these stars from $>1\,{\rm pc}$ by 
dynamical friction would exceed their inferred ages, unless the migration
rate is somehow accelerated.

In addition to their youth, these stars exhibit some peculiar dynamical
properties. One of them is strong clustering. 
The sources IRS~16 and IRS~13 were initially
thought to be single sources but later resolved into multiple components.
IRS~16 is a collection of young (3--7 Myr) He~{\sc i}
emission-line stars lying $1''$--$7''$ in projection 
from Sgr A$^\star$. 
They form a co-moving group
that is counter-rotating with respect to the Galaxy. IRS~13 is another
complex, composed of hot bright stars, about $3.6''$ south-west of
Sgr A$^\star$. In particular, IRS~13E is a 
compact, massive star cluster which is a few million years old 
\citep{MPSR04}. \cite{Genzeletal03} and \cite{Horrobinetal04} 
find that the young He~{\sc i} emission-line 
stars in the vicinity of Sgr A$^\star$ are
all concentrated in two disks, forming two co-moving (but not gravitationally
bound) populations. In addition to the clustering, stars 
within $\sim0.04\,{\rm pc}$ of 
Sgr A$^\star$, which are about 10\,Myr old, 
seem to have higher than normal eccentricities \citep{Schodel03}, but
this anomaly may be explained by selection effects \citep{Ghezetal04}.

One possible explanation for the presence of these young stars is that they are
not young but {\em rejuvenated\/}. A possible path to rejuvenation, stellar
mergers, is shown not to be viable by \cite{Genzeletal03}. Another
path, squeezars, heating of stars by close tidal encounters with the SBH
also fails to explain the observed population \citep{AM03}.

Aside from rejuvenation, 
another possible scenario, especially suggested by the observations of the
two counter rotating and coeval disks of young stars around the SBH, is
the infall and collision of two molecular 
clouds in the vicinity of the SBH. Such
a collision may provide the required densities for star formation.
\cite{Genzeletal03} suggest that the stars resulting from the two colliding clouds
will form two co-moving populations. However, since a collision leading to
high densities will be highly inelastic, it is not clear that 
two distinct populations will form. To our
knowledge, the collision and further evolution of dense clouds in the
vicinity of a SBH has never been studied in detail. 
So, it is hard to decide whether this is a viable scenario.

A possibility that we investigate in this paper is the inspiral of a star
cluster by dynamical friction, as suggested by \cite{Gerhard01}. \cite{PZMG03}
carried out $N$-body simulations to study this scenario. The clusters in
their simulations disrupted at $>1\,{\rm pc}$ from the GC, so cannot
explain the presence the young star population very close to Sgr A$^\star$.
\cite{KM03}, again using $N$-body simulations, concluded that, for a
cluster to reach within the central parsec, it must either be very
massive ($>10^6\,M_\odot$) or have formed near the GC (at $<5\,{\rm pc}$).
They found that, in both cases, a very high central density 
($\sim10^8\,M_\odot{\rm pc}^{-3}$) is required, and concluded that this
scenario is implausible.

\cite{HM03} suggested that the presence of an intermediate-mass black hole
(IMBH) can stabilize a cluster so that survival all the way to within the central
parsec can be achieved with much lower central densities. In addition,
they suggested that an IMBH--SBH binary can perturb the young stars into
radial orbits with small semi-major axes in a way similar to the Jupiter--Sun
system creating short period comets \citep{QTD90}.

\cite{KFM04} performed $N$-body simulations similar to those of
\cite{KM03} but with an additional IMBH embedded in the inspiraling cluster.
They found that an IMBH does decrease the requirement for a high central
density, but its mass must be about 10\% of the total cluster mass.
Since this is much larger than estimates of the collapsed
core mass in dynamical simulations \cite[0.1-0.2\%; see][]{PZM02,GFR04} 
they concluded that a realistic IMBH cannot help transport young stars
into the central parsec. However, after core collapse, 
the central object can continue to grow by colliding with the stars
that migrate to the center by relaxation. This growth continues
until the massive stars start evolving off the main sequence and lead
to cluster expansion by mass loss through winds and supernovae \citep{PZetal04,FGR04c}.

Using our dynamical Monte Carlo code, we have carried out simulations where
we form an IMBH through runaway collisions following core collapse during
the cluster's inspiral towards the GC. 
We discussed some initial results in a previous paper \citep{GR04a}. 
Here we present all our results from simulations of dense star clusters
near the GC using a more realistic Galactic mass distribution. We 
investigate the initial conditions required for these clusters to 
reach the GC within 3--10 Myr {\em and\/} 
undergo core collapse before disruption. In
\textsection 2 we describe our numerical technique for cluster simulations
and the inspiral mechanism. We present some semi-analytic calculations in
\textsection 3, and the results of our full simulations in \textsection 4. We
discuss these results and present our conclusions in \textsection 5.

\section{Initial Conditions and Numerical Technique}
\label{sec_num_tech}
For our simulations, we use a Monte Carlo technique, which provides an ideal
compromise in terms of speed and accuracy.  It has the star-by-star
nature of the $N$-body techniques but incorporates physical assumptions 
which lead to simplifications, 
reducing the computation time. As a result, we can carry out simulations of
systems with $N\gtrsim10^7$ stars in $\lesssim100$ CPU hours. For
details of our method, we refer to
our previous work (see \citealt{JRPZ00,JNR01} and references therein 
for basics, and \citealt{GFR04} for treatment of the 
realistic mass functions).
Here we only explain in detail the additions to our code for this paper.

\subsection{Initial conditions and Units}
For the initial mass function (IMF) we implemented a Salpeter distribution,
$dN/dm\propto m^{-2.35}$, with $m_{\rm min}=0.2\,M_\odot$ and 
$m_{\rm max}=120\, M_\odot$. Other choices of IMF, e.g., Miller-Scalo or Kroupa,
do not change the core collapse time, as long as $m_{\rm max}/m_{\rm min} \gtrsim 100$
\citep{GFR04}.  

For the initial structure of our clusters we chose King models 
\citep[see][Chapter 4]{BT87}.
This choice allows us to implement the initial tidal cutoff in a natural way.
In addition, the rate of evolution of the system, which is  determined
by the central relaxation time \citep{GFR04}, can be adjusted by a single
parameter, the dimensionless central potential $W_0$. For the sake of
simplicity we chose the tidal radius of the King model equal to the
Jacobi radius of the cluster. 
(see \textsection \ref{ssec_tidal_truncation})

Throughout this paper we use standard Fokker-Planck units \citep{Henon71b}:
$G = M_0 = -4E_0 = 1$, where $G$ is Newton's gravitational constant, 
$M_0$ is the initial total mass of the cluster, and $E_0$ is the initial
total energy.
The conversion to physical units is done by calculating the physical mass of
the system, and identifying the tidal radius of the King model used with the
Jacobi radius. Various length
and time scales of King models are given in Table 1 of \cite{GFR04}.

\subsection{Boundary Condition at the Center}
In the continuum limit, when a cluster undergoes core collapse,
the central density becomes infinite in a finite time. In addition to
being unphysical, high densities require small timesteps and render the
dynamical evolution of the system impossible to follow numerically.
When various physical processes and the finite radii of stars are taken into
account, the central density cannot
increase indefinitely. This can result from 
energy generation by 
``three-body binaries'' \citep[and references therein]{Giersz01}, ``burning''
of primordial binaries \citep[and references therein]{FGJR03} or physical collisions.
For the young dense clusters we consider here, we expect energy generation
from binary formation or binary burning to play a minor role, since for these
systems, an interaction with a hard binary is likely to lead to a merger
\citep{FCPZR04}.

The local collision time, i.e., the average time after which a
star has experienced one collision, is given by
\begin{equation}
 t_{\rm coll} \simeq 2.1\times 10^{12}\,{\rm yr}\,\frac{10^6\,{\rm pc}^{-3}}{n} 
\frac{\sigma_v}{30\,{\rm km\,s}^{-1}} \frac{R_\odot}{{\rm R}_\ast} \frac{M_\odot}
{{\rm M}_\ast},
\label{eq:t_rlx}
\end{equation}
where $\sigma_v$ is the
velocity dispersion, $n$ is the number density of stars, $R_\ast$ 
and $M_\ast$ are the radius and the mass of the star under consideration.
In the systems we consider in this paper, $t_{\rm coll}$ is larger
than the lifetime of the cluster except for the stars participating
in the core collapse. We used an approximation to exploit this property.
Rather than treating collisions 
explicitly as in \cite{FB02}, we introduced a simple boundary
condition near
the center of the cluster: when a star is part of the collapsing
core, we add its mass to a growing central point mass and remove it from the
simulation. 
This central point mass is then used only during the calculation of the 
cluster gravitational potential. We determine whether a star is part of the
collapsing core by monitoring its orbit's apocenter. We have chosen this
criterion rather than one based on pericenter or instantaneous position to
guarantee that the stars under consideration are restricted to a small region
near the center.
The threshold value we have chosen for the apocenter distance 
is $2\times10^{-4}$ 
in Fokker-Planck units. For our models, $t_{\rm coll}$ within this radius
is $\sim 10^4\,$yr, i.e., all stars restricted to this region
will rapidly undergo collisions. 
We determined the threshold value empirically. Choosing a large
value leads to removal of stars from the system before core collapse.
Choosing a very small value is also problematic because of the very
short collision times. If collisions are not taken into account, a few
massive stars sinking to the center of the cluster could easily reach
energies comparable to the total energy of the cluster and have
a substantial effect on the evolution of the system \citep[][and references
therein]{Henon75}. We found that threshold values in the range 
$5\times10^{-5}$--$3\times10^{-4}$ generally avoid these problems and
$2\times10^{-4}$ is a number suitable for all the clusters we
simulated.

We compared the rate of growth of the central mass in our simulations with
results from other Monte Carlo simulations where collisions between the stars
are treated more realistically \cite{FRB05,FGR05}.
We found that we typically underestimate the rate of growth slightly,
by 20 to 30\%. We have also compared the evolution of the average mass 
among the innermost
5\,000 stars and found that, in our simulations,
the increase in average mass slightly lags behind, but reaches the same final
value. This is probably related to the slower growth of the central mass 
in our code. It
is not possible to determine which results are more accurate as
these simulations cannot be repeated with more direct methods. In any case 
we do not expect this small difference to affect our results significantly.

\subsection{Tidal Truncation}
\label{ssec_tidal_truncation}
During the course of our simulations, we remove the stars that gain positive
energies because of interactions, or whose apocenter lies beyond the cluster's
tidal radius. This tidal radius depends on the mass distribution in the
GC region and the cluster's current position. 
The tidal radius of the
cluster can be estimated by using the following expression for the 
Jacobi radius
\begin{equation}
\label{eq_rJ}
r_J \simeq \left[\frac{m_J}{2 M(R)}\right]^{1/3} R \, ,
\end{equation}
where $m_J$ is the bound mass of the cluster, $R$ is the distance from the
GC, and $M(R)$ is the mass within a circular orbit at this
radius.
The latest estimate for the Galactic mass distribution near the central BH
is given by \cite{Genzeletal03}. They estimate the {\em stellar} mass density
as a broken power law,
\begin{equation}
\rho_\star(R) = 1.2\times10^6 \left(\frac{R}{R_b}\right)^{\alpha-3}
M_\sun\,{\rm pc}^{-3}\;,
\end{equation}
with $\alpha=1.63$ for $R\leqslant R_b$, $\alpha=1.0$ for $R>R_b$, and 
$R_b=0.38\,{\rm pc}$. 
This mass distribution is a self-consistent description that goes
all the way from 0.1 to about 10 pc, but much farther than that it may
underestimate the stellar density (Tal Alexander, private communication).
We derive the formulae for dynamical friction
on a point mass, resulting from a broken power law mass
density distribution and an additional central BH of mass 
$M_{\rm SBH} = 4\times10^6\,M_\sun$ \citep{MF01,Ghezetal03,Schodel03}, in
Appendix~\ref{app_1}. As a result of the 
dynamical friction, the distance of the cluster
from the GC changes continuously; we include the effect of this change by
adjusting $r_J$ appropriately in our simulations.

In principle, the treatment we use for dynamical friction is valid only
for point masses and extended clusters require a modification of this
treatment. However, the finite size affects only the Coulomb logarithm
in equation \ref{eq_fric_acc}, and, as long as the size of the cluster
($r_h$) is small with respect to the size of the region that contributes
to the dynamical friction, this leads only to a small decrease
in the drag on the cluster \citep[][\textsection~7.1]{BT87}.  
Since this condition is always satisfied in our simulations, we do not
expect our results to be affected by this approximation.
\cite{MPZ03} established numerically the validity of the formulation we
use by making comparisons to more exact $N$-body simulations.

\section{Semi-analytical calculations}
\begin{figure}
\resizebox{\hsize}{!}{\includegraphics[clip]{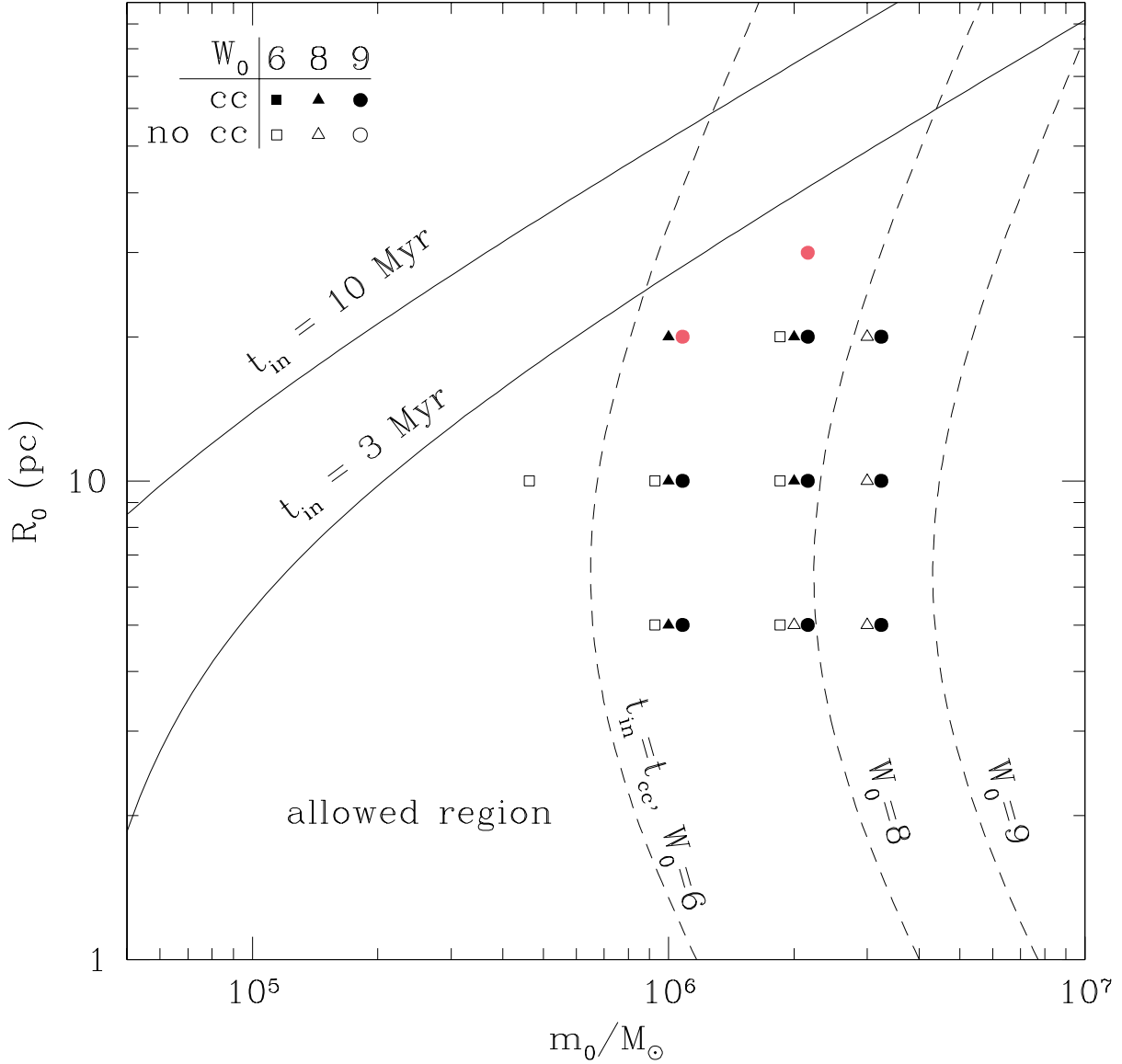}}
\caption{
Region of $R_0$-$m_0$ space that allows fast inspiral (denoted by solid
lines) and core collapse happening during inspiral (for various King models;
denoted by dashed lines), in the point-mass approximation. Results of various simulations are shown with symbols of different shapes, corresponding to
different initial structure; full symbols are for systems that experienced
core collapse and open symbols correspond to systems that disrupted before
core collapse. All  
models shown have $t_{\rm dis}<3\,{\rm Myr}$ except for the ones with gray symbols
(colored red in the electronic edition), which have 
$3\,{\rm Myr}<t_{\rm dis}<4\,{\rm Myr}$
\label{fig_const}
}
\end{figure}

For a point mass, the time required to reach the GC from an initial distance
$R_0$ can be calculated by integrating equations (\ref{eq_dim_dec}) and
(\ref{eq_spi_in2}). Since equation (\ref{eq_dim_dec}) has a logarithmic
singularity near the origin, for the calculations below, we stop the
integration at $R=0.01\,{\rm pc}$. Most of the mass loss from the
cluster takes place near the center, so neglecting the mass loss provides a
reasonable first estimate for the inspiral time, $t_{\rm in}$.  Possible 
upper limits for $t_{\rm in}$ are 3 or 10 Myr, which are about the lifetimes of
the brightest IR stars observed near the GC. By requiring the cluster to
undergo core collapse before reaching the GC, we can also obtain a lower
limit for $t_{\rm in}$. 
This can be transformed into a condition on the initial structure of
the cluster using $t_{\rm cc}\simeq 0.15\,t_{\rm rc}(0)$ \citep{GFR04},
where $t_{\rm rc}(0)$ is the initial central relaxation time.
We illustrate these constraints in Figure \ref{fig_const}. The solid
lines correspond to the upper limits on $t_{\rm in}$ from 
the stellar evolution timescale.
Above these lines it takes more than the denoted time (3 or 10 Myr)
to reach 0.01 pc for a point mass.
The bow-like dashed lines correspond to lower limits 
for various King models. To the
right of these lines, the inspiral time for a point mass is shorter than the
core collapse time, hence runaway collisions cannot happen unless the system
is initially collisional. 
For these calculations we have computed the timescales for the King models
by assuming their tidal radius is initially equal to their Jacobi radius
given in Equation \ref{eq_rJ}. In this figure we also show results of some
of our simulations (see next section).

\begin{figure}
\resizebox{\hsize}{!}{\includegraphics[clip]{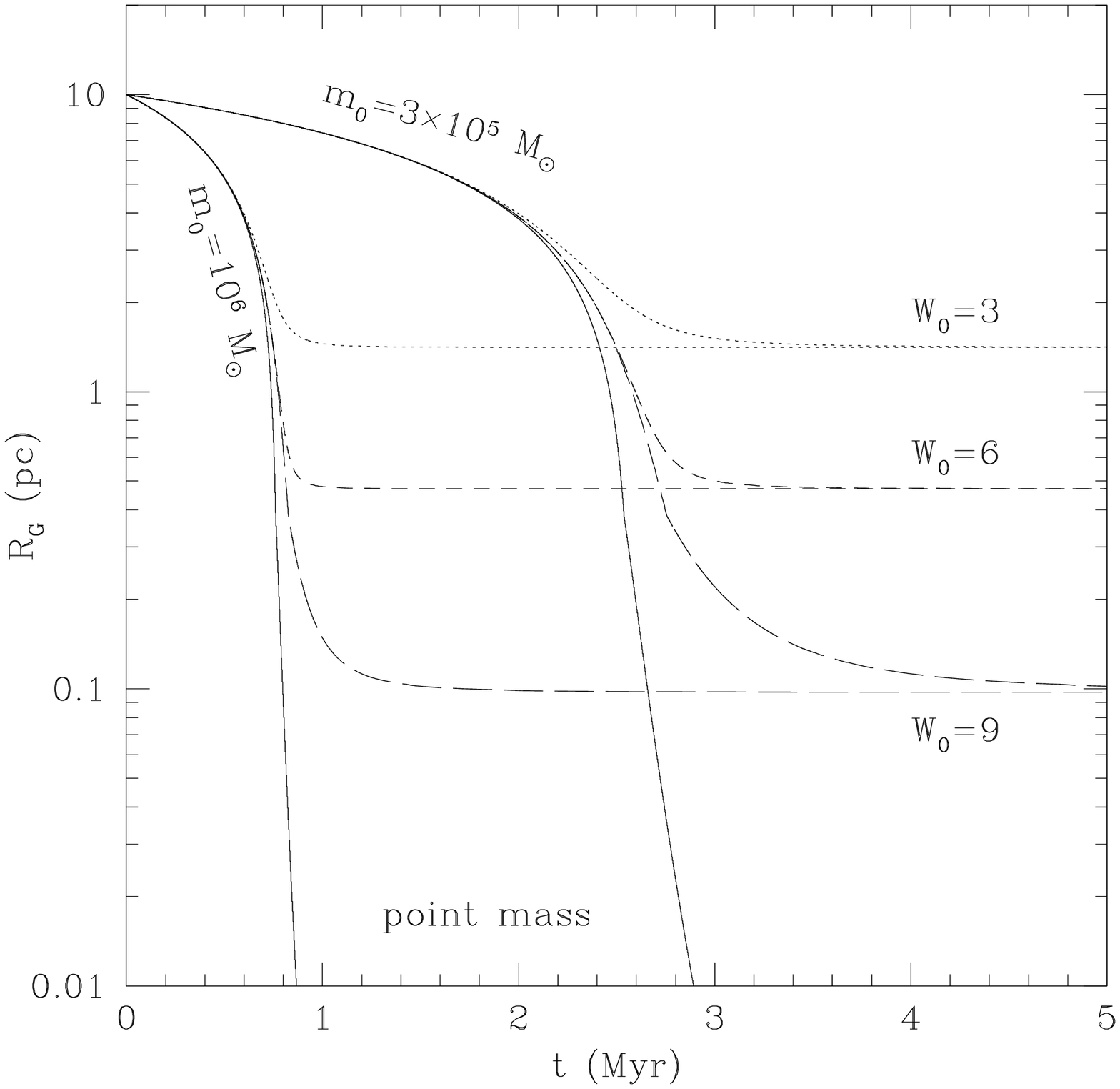}}
\caption{
Rate of inspiral for point-mass and King models. The two families
of curves correspond to initial masses $m_0 = 3\times 10^5 {M_\odot}$ and $m_0=10^6 M_\sun$. For each initial mass, results are shown for point-mass (solid line) and King models with $W_0=9$ (long-dashed line), $W_0=6$ (dashed line), and $W_0=3$ (dotted line).
\label{fig_R_vs_t}
}
\end{figure}

The next step in approximation is using a realistic structure for the cluster
but neglecting the changes in structure during the inspiral.
Since the structure changes on a
relaxation timescale and in the outer parts of the cluster the relaxation time
is much longer than the inspiral time, this approximation provides a good
estimate for the mass loss during the inspiral, at least until close to
disruption. We have computed the rate of inspiral of 
different King models for different initial distances from the GC and initial
cluster masses. We present our results for initial distance
$R_0=10\,{\rm pc}$ in Figure~\ref{fig_R_vs_t}.
During the inspiral we calculate the change in bound mass as follows. First
we calculate the tidal radius by using the current mass and distance from
the GC by equation~(\ref{eq_rJ}). From this and the structure of the given
King model a new bound mass is calculated. This procedure is repeated
until the bound mass converges to a value. Failure of convergence marks
the disruption. 

\begin{figure}
\resizebox{\hsize}{!}{\includegraphics[clip]{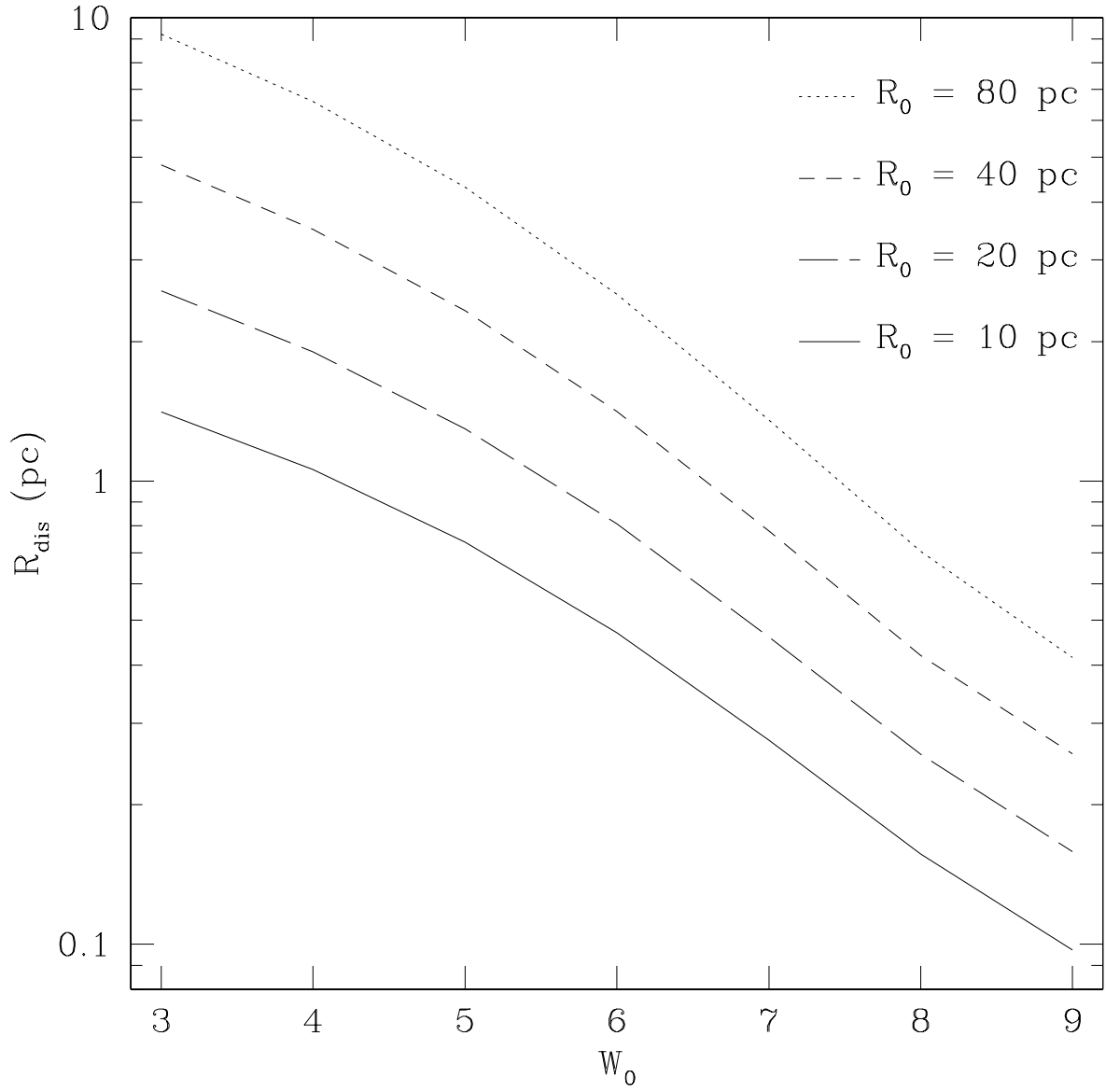}}
\caption{
Dependence of disruption radius $R_{\rm dis}$ on initial concentration and
initial distance.
\label{fig_Rdis_vs_w0}
}
\end{figure}

In Figures \ref{fig_R_vs_t} and \ref{fig_Rdis_vs_w0} we show the results of our
calculations for extended clusters. Unlike a point-mass, 
an extended cluster disrupts at a
finite distance from the GC, $R_{\rm dis}$. 
This distance decreases as more of the cluster's mass is concentrated near its
center, but is independent of the total mass in the cluster.
The inspiral
time, on the other hand, has a strong dependence on the total mass. 
The independence of $R_{\rm dis}$ from total mass can be understood as
follows. For a given $W_0$, the disruption happens when the tidal radius
reaches a specific fraction of the initial Jacobi radius. The relation
between this value and the distance from the GC is independent of the
initial cluster mass, since more massive clusters start with a 
proportionately larger initial Jacobi radius.
This is also the main reason for the dependence of $R_{\rm dis}$ on initial
distance. Note that, in principle, a cluster can underfill
its Jacobi radius at formation, i.e., clusters formed at large distances can have the same size as the clusters formed close to the GC.
The disruption radii we obtain in this approximation are upper
limits since mass segregation, the formation of a cusp, and the subsequent 
IMBH will all increase the mass concentration towards the center.

\begin{figure}
\resizebox{\hsize}{!}{\includegraphics[clip]{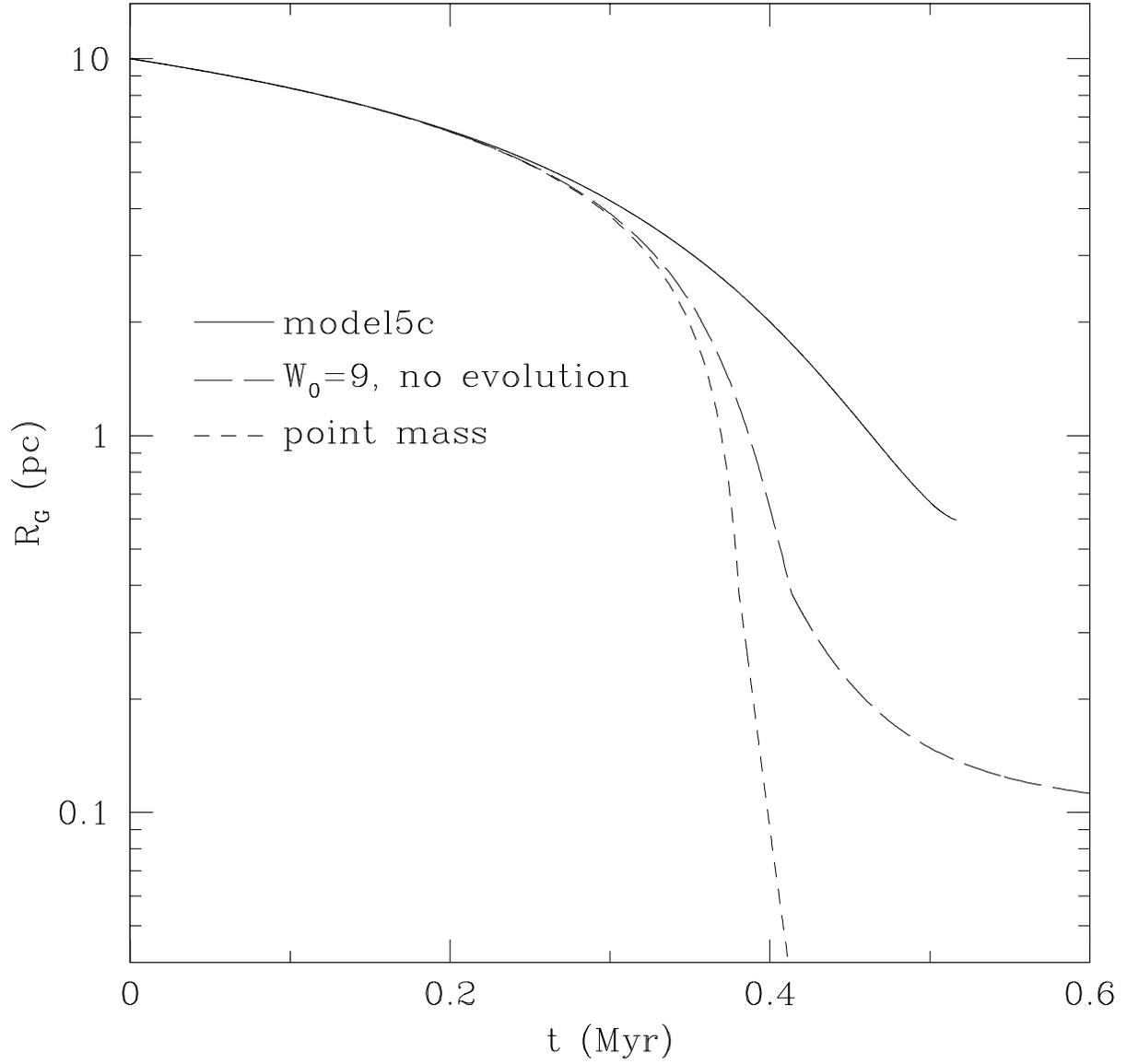}}
\caption{
Comparison of a full simulation with an extended static model and the 
point-mass
approximation.
\label{fig_comp_three}
}
\end{figure}

These calculations show that for a wide region of $R_0 {\rm -} m_0$ parameter
space it is plausible that a cluster can reach to within 
a fraction of a parsec
of the GC in a few million years. A high initial concentration ($W_0\gtrsim8$) 
is required for the cluster not to disrupt too far from the GC, as well
as to guarantee rapid core collapse. More
detailed simulations are required to verify and establish these findings,
and to obtain the demographics of the stars that reach the GC.
To illustrate this point, we present a comparison of point-mass and extended
cluster models with a full simulation (Model5c, see next section) in 
Figure~\ref{fig_comp_three}. All calculations presented in this figure
start with the same initial mass and initial distance from the GC. This comparison
shows that the effects of dynamical evolution, namely the enhanced mass loss due
to expansion, can be quite important. Note that, for clusters with lower
initial concentrations, the relaxation will be slower and the difference between
full simulation and the extended static model will be smaller. However the
deviations from the point-mass approximation will be larger.

\section{Results of simulations}
\subsection{Overview of results}
We have performed simulations starting with King models of dimensionless
central potentials $W_0=6, 8,\,{\rm and}\,9$ at
a variety of distances and initial masses. We present our results in Table~\ref{tab_sim_res}.
We ended our simulations when the number of stars in the cluster dropped to
0.5\% of its initial value. We defined this as the point of disruption and
denoted the time to reach this point and distance from GC with $t_{\rm dis}$ and $R_{\rm dis}$,
respectively. Allowing the cluster to evolve beyond this point will not
change our results since, at this point, the mass is so low that 
dynamical friction is no longer
effective, and the distance from the GC remains constant. In addition, as a 
result of the low number of stars, relaxation and the consequent evaporation
of the cluster will be very rapid.
Most of the clusters with high initial concentration went into core collapse before disruption and the central point mass described in \textsection \ref{sec_num_tech} grew during the inspiral. This accumulated mass is indicated in column 7 of our table. For less dense clusters, the inspiral is faster than the core collapse and in those cases no central point mass is grown. 
We did not observe significant post-core-collapse expansion in our
simulations. The growth of the central mass indicates that segregation of
massive stars continue beyond the core collapse.

\begin{deluxetable}{lrrrrrr}
\tabletypesize{\scriptsize}
\tablecaption{Results of Simulations
\label{tab_sim_res}}
\tablewidth{0pt}
\tablehead{
\colhead{$\tbfrac{\rm model}{\rm ID}$} & 
\colhead{$\tbfrac{m_0}{(M_\sun)}$} & 
\colhead{$\tbfrac{R_0}{\rm (pc)}$} & 
\colhead{$W_0$} & 
\colhead{$\tbfrac{R_{\rm dis}}{\rm (pc)}$} & 
\colhead{$\tbfrac{t_{\rm dis}}{\rm (Myr)}$} & 
\colhead{$\tbfrac{m_{\rm cen}}{(M_\sun)}$}
}
\startdata
1a\dotfill  & $3\times10^6$ 	& 20 & 8  & 1.04 & 1.04 & - \\ 
2a\dotfill  & $3\times10^6$ 	& 10 & 8  & 0.35 & 0.31 & - \\  
3a\dotfill  & $3\times10^6$ 	&  5 & 8  & 0.22 & 0.12 & - \\  
4a\dotfill  & $2\times10^6$ 	& 20 & 8  & 1.58 & 1.59 & 3000 \\  
5a\dotfill  & $2\times10^6$ 	& 10 & 8  & 0.88 & 0.46 & 580 \\  
6a\dotfill  & $2\times10^6$ 	&  5 & 8  & 0.34 & 0.17 & - \\
7a\dotfill  & $10^6$ 		& 20 & 8  & 1.89 & 3.16 & 3600 \\  
8a\dotfill  & $10^6$ 		& 10 & 8  & 0.98 & 0.94 & 2500 \\  
9a\dotfill  & $10^6$		&  5 & 8  & 0.47 & 0.32 & 2000 \\ 
\hline  
4b\dotfill  & $2\times10^6$ 	& 20 & 6  & 1.44 & 3.00 & - \\  
5b\dotfill  & $2\times10^6$ 	& 10 & 6  & 0.42 & 0.85 & - \\  
6b\dotfill  & $2\times10^6$ 	&  5 & 6  & 0.24 & 0.31 & - \\  
7b\dotfill  & $10^6$ 		& 20 & 6  & 2.09 & 3.08 & - \\  
8b\dotfill  & $10^6$ 		& 10 & 6  & 0.92 & 0.88 & - \\  
9b\dotfill  & $10^6$		&  5 & 6  & 0.42 & 0.30 & - \\  
10b\dotfill & $5\times10^5$	& 20 & 6  & 4.59 & 2.96 & 560 \\  
11b\dotfill & $5\times10^5$	& 10 & 6  & 1.55 & 0.86 & - \\  
12b\dotfill & $3\times10^5$	& 20 & 6  & 4.44 & 3.00 & 1200 \\  
13b\dotfill & $3\times10^5$	& 10 & 6  & 2.31 & 0.80 & - \\ 
\hline  
1c\dotfill  & $3\times10^6$ 	& 20 & 9  & 1.23 & 1.10 & 5500 \\ 
2c\dotfill  & $3\times10^6$ 	& 10 & 9  & 0.56 & 0.35 & 4500 \\  
3c\dotfill  & $3\times10^6$ 	&  5 & 9  & 0.28 & 0.14 & 3000 \\  
4c\dotfill  & $2\times10^6$ 	& 20 & 9  & 1.20 & 1.70 & 6000 \\  
5c\dotfill  & $2\times10^6$ 	& 10 & 9  & 0.60 & 0.52 & 4900 \\  
6c\dotfill  & $2\times10^6$ 	&  5 & 9  & 0.33 & 0.20 & 4200 \\  
7c\dotfill  & $10^6$ 		& 20 & 9  & 1.89 & 3.23 & 4700 \\  
8c\dotfill  & $10^6$ 		& 10 & 9  & 0.76 & 1.04 & 3500 \\  
9c\dotfill  & $10^6$		&  5 & 9  & 0.43 & 0.34 & 3600 \\  
10c\dotfill & $5\times10^5$	& 20 & 9  & 3.14 & 6.50 & 2900 \\  
11c\dotfill & $5\times10^5$	& 10 & 9  & 1.26 & 1.95 & 2700 \\  
12c\dotfill & $3\times10^5$	& 20 & 9  & 4.18 & 11.0 & 1900 \\  
13c\dotfill & $3\times10^5$	& 10 & 9  & 1.91 & 3.04 & 2000 \\  
17c\dotfill & $2\times10^6$	& 30 & 9  & 1.96 & 3.54 & 6800 \\  
18c\dotfill & $2\times10^6$	& 60 & 9  & 5.17 & 13.8 & 6800 \\  
19c\dotfill & $2\times10^6$	& 50 & 9  & 3.81 & 9.63 & 6600
\enddata
\tablecomments{
The first column is the model ID. The following three columns indicate the 
initial conditions of the cluster: mass in solar masses, initial distance
from the GC in parsecs and the dimensionless central potential. The last
three columns are the results of the simulations: the disruption distance
in parsecs, the time it takes for disruption in Myr, and the accumulated
mass at the center of the cluster in solar masses. If the cluster disrupts before
going into core collapse, no mass is accumulated at the center. In these cases
a  hyphen is used in the last column.
}
\end{deluxetable}

An overview of the results shows that neglecting the changes in the
structure of the cluster during its inspiral provides a good estimate for
the time to disruption, $t_{\rm dis}$. However, this is not true for 
$R_{\rm dis}$, which is dependent
on the total initial mass for an evolving cluster. This dependence is a result
of the interplay between inspiral time and relaxation time. For a massive cluster
containing a large number of stars, the inspiral time is short but the relaxation
time is long. Hence only the central part of the cluster has time to evolve. For
a less massive cluster the outer parts have time to evolve and expand. As a result,
the mass loss is enhanced and the disruption happens at a further distance from 
the GC. 

In Figure \ref{fig_const} we plot the results of some our simulations to
compare with point-mass estimates. We have only plotted models where disruption
distance from the GC was small. For models started with $R_0\le10\,{\rm pc}$,
only the ones within 1\,pc of the GC; and for the others only the
ones disrupted within 2\,pc of GC are plotted.
We have used black symbols for models with $t_{\rm dis}<3\,{\rm Myr}$, 
and gray symbols (colored red in the online version) otherwise. 
We find that inspiral and/or disruption time
in simulations exceed the estimated inspiral time from point-mass approximation.
Similarly, the disruption distance is always larger
in full simulations than static cluster approximations. Finally, sometimes
clusters disrupt without going into core collapse in simulations where
the point-mass 
approximation predicts core collapse times shorter than inspiral
times. These results show that semi-analytical approximations provide
necessary but not sufficient conditions for our scenario to work. Increasing
the realism by taking the tidal mass loss of the cluster and its expansion
into account cannot be compensated by the increase in central density of the
cluster, and leads to disruption happening earlier and/or further from
the GC than indicated by the semi-analytical methods. It is not possible
to resolve the core-collapse criterion, indicated by dashed lines in
Figure \ref{fig_const}, for $W_0=6, 9$ models as is done for $W_0=8$ case.
For $W_0=6$, the boundary is at low initial mass which correspond to low
number of stars. Such clusters have short relaxation times and hence
expand and disrupt rapidly before going into core collapse unless they
start far from the GC. In contrast, the boundary for $W_0=9$ lies at
high initial mass and large number of particles, making the simulations
of such clusters computationally impractical.

Another trend that can be seen in Table~\ref{tab_sim_res} 
is that more mass is accumulated at the
center for clusters that start at larger distances from the GC. This is simply
because there is more time available for the mass buildup. This does not necessarily mean that a heavier IMBH will be formed in these systems, as the fate of a very massive star being bombarded by other stars is not certain. Also, 
being the total available mass, this is strictly 
an {\em upper limit} to rather than an estimate of the IMBH mass.

\begin{figure}
\resizebox{\hsize}{!}{\includegraphics[clip]{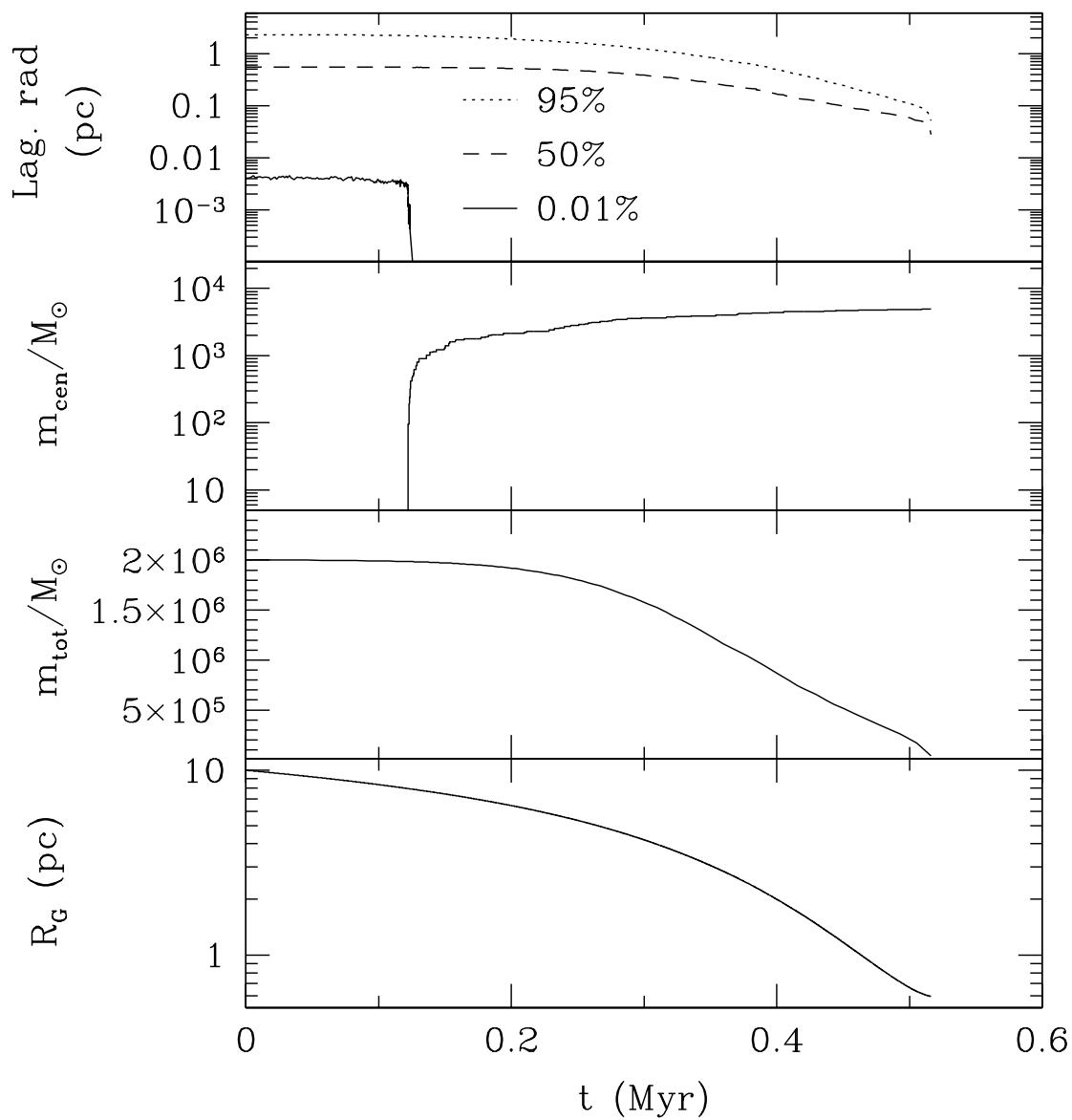}}
\caption{
Evolution of Model 5c (see Table \ref{tab_sim_res} for initial parameters). The top panel shows 0.01\%, 50\% and 95\% Lagrange radii. The second panel is the growth
of the central mass, starting at the core collapse. The third panel shows the 
evolution of the bound mass of the cluster and the bottom panel is the distance of 
the cluster from the GC.
\label{fig_model5c_evol}
}
\end{figure}

In the rest of this section we present an analysis of the evolution of
our Model 5c (see Figure \ref{fig_model5c_evol}), which started as a $W_0=9$ King model, with mass $m_0=2\times10^6\,M_\sun$
at distance $R_0=10\,{\rm pc}$. 
This cluster went into core collapse very rapidly ($\gtrsim 10^5$ years)
and reached the GC in about $5\times 10^5$ years. After core collapse the central mass grew to a value $M_{\rm cen}\simeq 5000 M_\odot$, which is consistent with the estimate, $M_{\rm cen}\simeq 0.002 M_{\rm total}$, by \cite{GFR04}.
The disruption of the cluster took place at $R_{\rm dis}=0.6\,{\rm pc}$ from the GC.

\subsection{Stars Stripped from the Cluster}
\begin{figure}
\resizebox{\hsize}{!}{\includegraphics[clip]{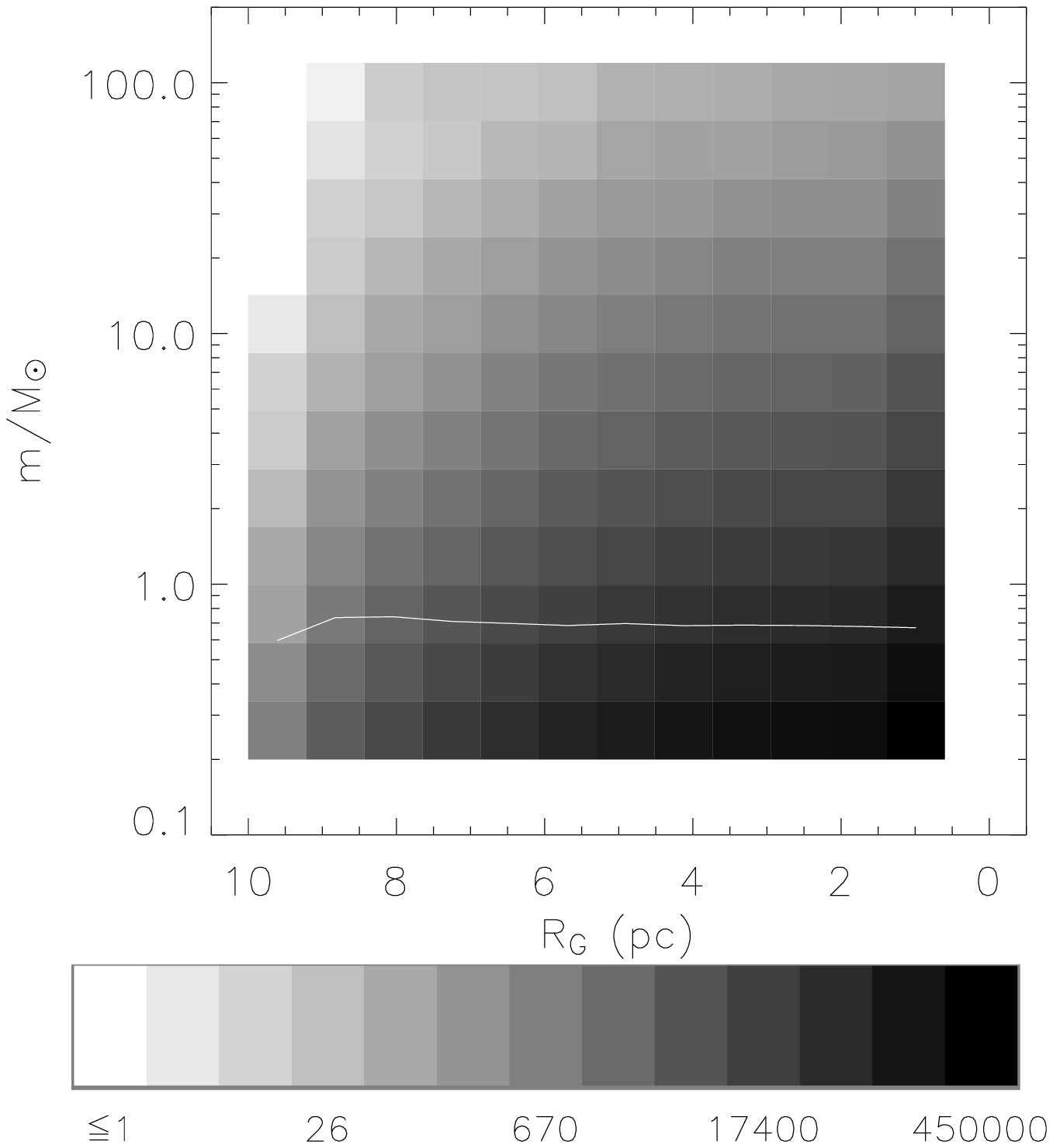}}
\caption{
Mass and distance distribution of stars that leave the cluster during inspiral.
The grey scale for the number of stars is logarithmic except for the 
$\leqslant 1$ bin; the average mass in each distance bin is shown by a white line. 
\label{fig_R_vs_m0}
}
\end{figure}

\begin{figure}
\resizebox{\hsize}{!}{\includegraphics[clip]{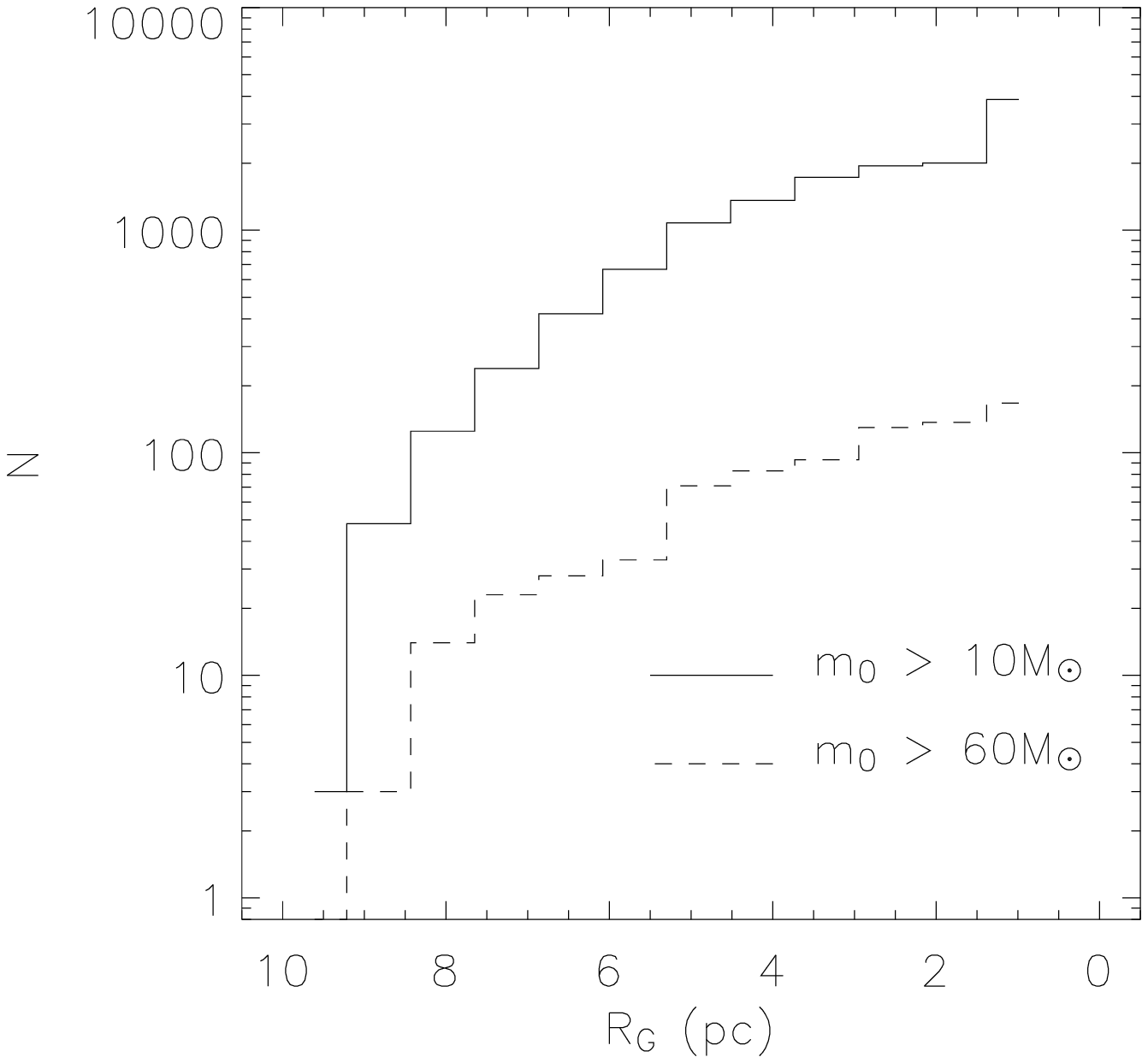}}
\caption{Number of massive stars that left the cluster at different distances from the GC. The solid and dashed lines indicate the stars whose initial
mass is larger than $10 M_\odot$ and  $60 M_\odot$, respectively.
\label{fig_R_vs_N}
}
\end{figure}

During the inspiral, stars become unbound from 
the cluster as a result of the shrinking Jacobi
radius. We present the mass distribution of these stars with respect to
their distance from the GC in Figure~\ref{fig_R_vs_m0}, and the number of massive stars that leave the cluster within each distance bin in Figure~\ref{fig_R_vs_N}.
In these figures it is 
seen that the massive stars leave the cluster predominantly near disruption. 
This is because of mass segregation, but also because
most stars leave the cluster near disruption and there are relatively
few massive stars. As expected, the average mass of the stars that leave 
the cluster (indicated by the white line in Figure~\ref{fig_R_vs_m0}) slightly decreases throughout the inspiral (see also Fig.~7 of \citealt{GFR04}). 
As the cluster evolves, massive stars sink towards the
cluster center, leaving behind less massive stars which are preferentially
removed. Our choice of initial conditions (requirement of core collapse
before the end of inspiral) allows massive stars to segregate faster than
the shrinking of the Jacobi radius because of the
inspiral, so most of them remain bound
throughout.

\begin{figure}
\resizebox{\hsize}{!}{\includegraphics[clip]{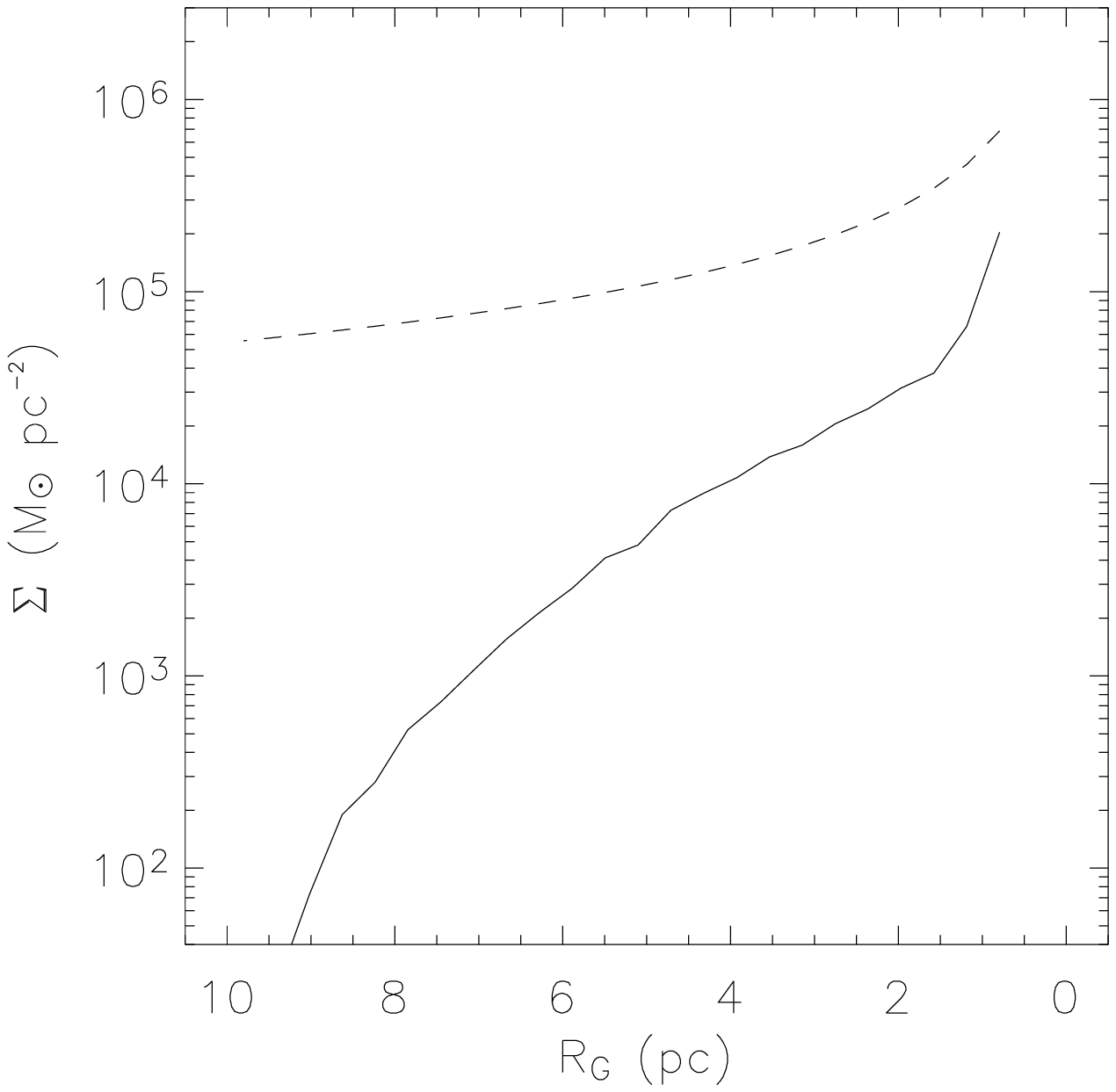}}
\caption{
Surface density of stars that leave the cluster (solid line) compared with the Galactic mass density used in calculations (dashed line). Note that for the cluster, the surface density is calculated using the initial masses of 
the stars. 
\label{fig_surf_den_prof}
}
\end{figure}

We present the surface density of initial masses of the stars that leave 
the cluster and compare this with the assumed mass density profile of the
Galaxy in Figure \ref{fig_surf_den_prof}. Throughout the inspiral, the contribution to the background surface density is small. 
The only detectable signature of a cluster inspiral will be the stars
left behind that are significantly  more massive and hence brighter
than the stars of the Galactic background distribution.

\subsection{The Cusp Retained at Disruption and IMBH}
Near disruption, the cluster forms a power-law cusp, $\rho\propto r^{-\alpha}$, around the central point mass. We estimated the power-law
exponent to be 
$1.35 < \alpha < 1.60$, which is compatible with the results of 
theoretical calculations \citep{BW77} and $N$-body simulations \citep{PMS04,Baumgardtetal04b,Baumgardtetal04c} which yield a value $\simeq 1.55$.

\begin{figure}
\resizebox{\hsize}{!}{\includegraphics[clip]{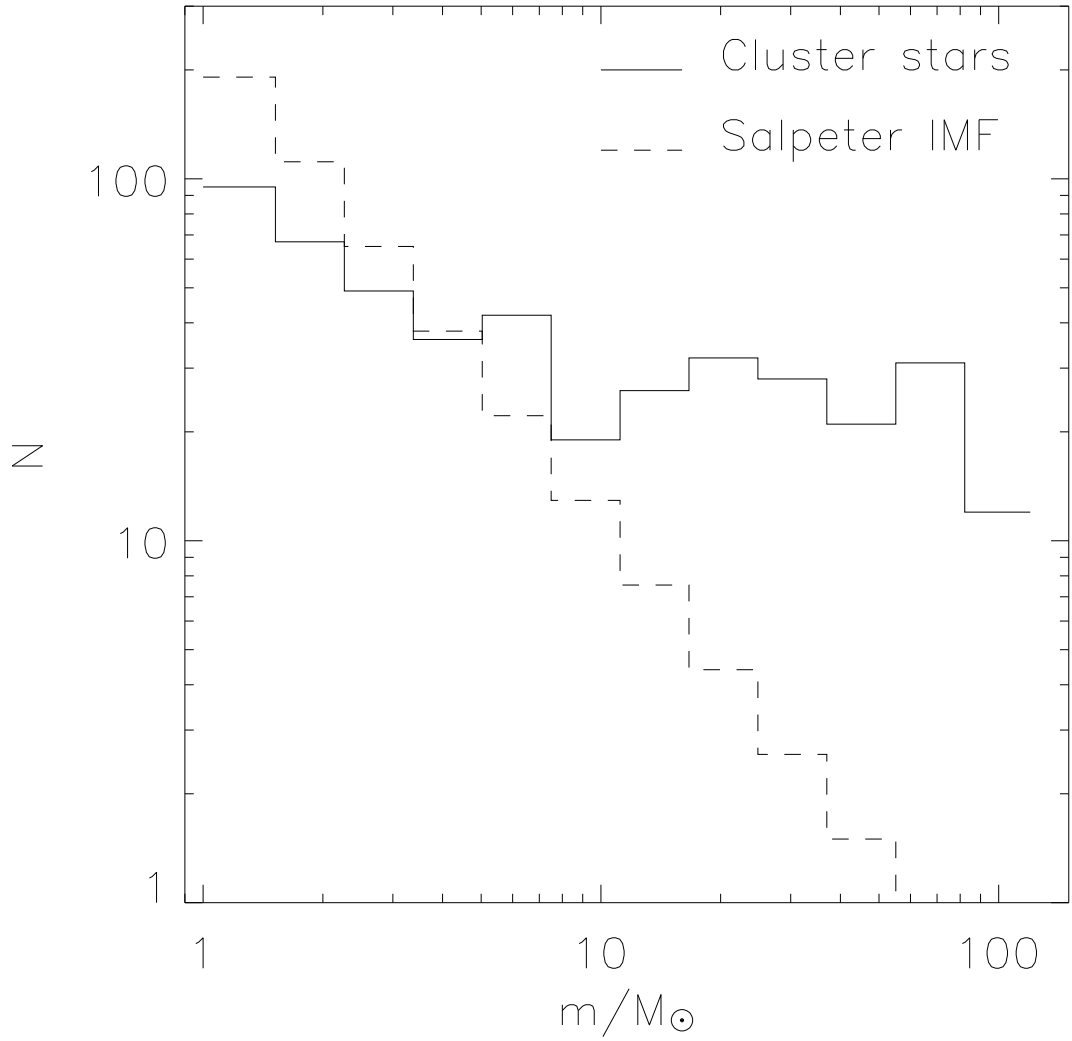}}
\caption{
Comparison of the mass function for the innermost 2000 stars at cluster
disruption (solid line) with the Salpeter IMF (dashed line; normalized to the same number of stars). The heavy part of the mass function at disruption
is much more populated because of mass segregation.
\label{fig_mass_functions}
}
\end{figure}

The mass function of the stars in this cusp is quite different from the 
Salpeter IMF which we adopted at $t=0$. The heavy part of the
mass function is more 
populated because of mass segregation, even though the
massive stars are preferentially removed via mergers with 
the central point mass.
We show a comparison of the mass function for the innermost 2000 stars with 
the Salpeter IMF in Figure \ref{fig_mass_functions}. 

At disruption, the density of stars from the cluster exceeds the 
density of background stars. The (3-d)
velocity dispersion at disruption is about 90 km\,s$^{-1}$. 
Since this is much
smaller than the velocities near the GC, the stars would stay together and
be observed as a clump, similar to the Central cluster \citep{Figer04},
or the rotating disk-like structures observed by \cite{Genzeletal03} and
\cite{Horrobinetal04}. Additionally, the stars which lie very close to
the IMBH will remain bound to it longer than the
stars lying outside, since these stars will be on Keplerian orbits and
their relaxation time will be longer, preventing the expansion of the
cluster. Such a collection of stars may exhibit a structure similar to
that of the IRS~13E cluster observed by \cite{MPSR04}.

It is not possible to follow reliably the evolution of the cluster near
disruption with our method, for several reasons. 
When the number of stars decreases to a very small value 
($N\sim1000$, see \citealt{Henon73}, \textsection 2.4.1), 
the relaxation time becomes comparable to the dynamical time, making
it necessary to take the effect of large-angle scatterings into account.
This would be possible in principle, by incorporating these
interactions into the Monte Carlo scheme in a way similar to collisions
\citep{FB02}, if the rate of these events were low and the relaxation
was still the dominant process for the evolution of the cluster. However,
in this regime, the dynamical evolution, 
in particular the evaporation, of the cluster
will not progress on a relaxation timescale and will depend sensitively 
on these large-angle scatterings. 
In addition, as the stars leave
the cluster, the potential will get shallower. As a result, the restoring
force on the central black hole, which keeps it near the cluster center,
will decrease and its wandering will increase (which we do not model).

\section{Discussion}
The central parsec of our Galaxy hosts many young stars whose presence
there poses a problem because of the unfavorable conditions for their
formation. The range of age estimates and the peculiar dynamical properties
(disk-like structures, clustering, possibility of larger than
normal eccentricities) suggest that there is more than one mechanism at
play. In this paper we consider one of the possibilities.

We investigated whether it is possible to explain the presence of
the young massive stars near the GC by bringing them there as
members of a cluster. At the same time, we want this cluster to
form an IMBH, which can then make the stars in its vicinity migrate to
inner orbits very near Sgr A$^\star$.
This scenario requires: (1) rapid inspiral of the cluster, 
$t_{\rm in} \lesssim 3\,{\rm Myr}$; (2) the disruption to take place
close to the GC, $R_{\rm dis} \lesssim 1\,{\rm pc}$; (3) the cluster
to undergo core collapse during the inspiral to form an IMBH.
Our work extends that of \cite{Gerhard01} and \cite{MPZ03}, 
who carried out semi-analytical calculations, and of 
\cite{PZMG03}, \cite{KM03} and \cite{KFM04}, who used $N$-body simulations
of dense clusters spiraling into the GC.

To determine the initial conditions suitable for this scenario, we 
carried out semi-analytical calculations and dynamical 
Monte Carlo simulations.
A comparison of the results obtained by these methods showed that
the change in the cluster structure has a 
considerable influence on the disruption 
distance. Following the evolution of the cluster also
allowed us to draw conclusions about the demographics of the stars
left behind and brought into the central parsec.
The expansion of the outer parts of the cluster because of the relaxation
enhances the mass loss and leads to earlier disruption.

Our simulations
show that for clusters with $R_0\gtrsim10\,{\rm pc}$, 
an initial mass of at least $10^6\,M_\odot$ is necessary. This is 
in agreement with the findings of \cite{KM03}.
We also found that high concentrations
($W_0\gtrsim8$) are required to undergo
core collapse during the inspiral. 
However, if the core collapse requirement is relaxed,
clusters with moderate concentrations ($W_0\sim6$) can also survive
the inspiral down to $R_{\rm dis}<1\,{\rm pc}$, starting with
$R_0\gtrsim10\,{\rm pc}$ and $m_0\gtrsim10^6\,M_\odot$. This is because
core collapse requires significant relaxation and clusters with low
concentration cannot survive the accompanying expansion.

The density of the stars that leave the cluster during the inspiral
is generally small compared to the Galactic background density, 
except near disruption. Almost all the massive stars that leave
the cluster during the inspiral do so near disruption. As a result
of the mass segregation, most of the cluster mass close to the IMBH is
in heavy ($m>10\,M_\odot$) stars. Upon disruption, these stars will
end up on orbits close to the IMBH. As a result of their proximity,
they can then undergo strong interactions with the IMBH and possibly get 
scattered into orbits closer to the GC. If such clusters harboring
IMBHs regularly form in the GC region, the central parsec may be
hosting more than one IMBH and this process can be realized by 
participation of multiple IMBHs. The age spread of the young stars in
the central parsec ($\sim$3--10\,Myr) implies that 
there has been more than
one instance of star formation in the recent history of the GC region.

Although the discussion in this paper has focused on our Galactic
center, these ideas also have important consequences for other galaxies and for extragalactic astrophysics.
The direct injection into the center of a galaxy of many IMBHs produced
by collisional runaways in nearby young star clusters provides a
potential new channel for building up
the mass of a central SBH through massive BH mergers \citep{PZM02}. 
In contrast, minor mergers of
galaxies are unlikely to produce massive BH mergers, as the smaller BH
will rarely experience enough dynamical friction to spiral in all the way to
the center of
the more massive galaxy \citep{VHM03}. Our scenario has important
consequences for LISA, the Laser-Interferometer Space Antenna, since the 
inspiral of an IMBH
into a SBH provides an opportunity for direct
study of strong field gravity and for testing general
relativity \citep{CH04,CT02,Phinney03}.
Although the SBHs found in bright quasars and many nearby galactic
nuclei are thought
to have grown mainly by gas accretion \citep[e.g.][]{FI99,HNR98,Richstone04,Soltan82},
current models suggest that LISA will probe most efficiently a
cosmological massive BH population of lower mass, which is largely
undetected \citep{Menou03}. LISA will measure their masses with exquisite
accuracy, and their mass spectrum will constrain formation scenarios
for high-redshift, low-mass galaxies and, more generally, hierarchical
models of galaxy formation \citep[e.g.][]{HK02,HH03,SHMV04,VHM03}.

The main difficulty with our scenario is the requirement of large
initial cluster masses for rapid inspiral from $R_0\gtrsim10\,{\rm pc}$.
A massive ($m_0>10^5\,M_\odot$) cluster with a Salpeter IMF will contain a
larger number of massive stars than currently observed at the GC region, both
within and outside the central parsec. In particular, outside the central
parsec, only a few young stars have been observed \citep{Coteraetal99}, whereas we
predict a larger number of these stars would have left the cluster during
inspiral. There are multiple ways to resolve this problem, but it is not
possible to determine by observation which, if any, of these mechanisms is
most important.  One of them is the formation of a cluster with a steeper IMF at the
higher mass end, hence suppressing the number of massive stars. The upper
mass cutoff we have chosen ($m_{\rm max}=120\,M_\odot$) is equivalent to
introducing a steep IMF \citep{WK04}, but even more conservative choices
are plausible (P.\ Kroupa, private communication). 
The requirement of a large initial
mass is somewhat relaxed if the cluster is initially on an eccentric orbit
\citep{KFM04}. In this paper, we only considered circular orbits, so the
mass requirements we find can be seen as upper limits. The problem of many
young stars leaving the cluster during inspiral would be largely avoided
by the presence of {\em initial\/} mass segregation, which is supported both by
observations and theoretical arguments \citep{ML96,BD98,RM98,dGGJ03}.
If the most massive members of the cluster start their life closer to the
center, they will have less of a chance to leave the cluster outside the
central parsec.  We also note that the lack of observed young stars outside the central
parsec could also result from higher extinction in this region
(F.\ Yusef-Zadeh, private communication).

The initial mass function of the Arches cluster is thought
to be significantly shallower than the Salpeter mass function
\citep{Figeretal99,KFLM00}. If a cluster is formed with a such an IMF,
the central mass will grow faster and larger, since there are more massive
stars.  This may decrease the required total
mass of the cluster, at the expense of increasing the fraction of mass in
massive stars, since a larger central object can support a cluster more
efficiently. We carried out a few simulations with shallower IMFs, power-law
mass functions with $\alpha=2.00 and 1.75$. To make a fair comparison, we
kept the total mass and the total number of stars in these cluster the same
as in our Salpeter IMF models, and hence modified the minimum mass in the
cluster. This way, the relaxation time is kept constant, and the rate of
expansion is expected to be similar for different IMFs. We found that, even
though central masses up to twice as large were grown, the changes in the
disruption distance were very small in these simulations.  We conclude
that growing a large enough central mass to have significant impact on
this scenario \citep[~10\% of initial total cluster mass][]{KFM04} is not
possible by a modification of the IMF.  Note that the mass function of the
innermost 2000 stars at cluster disruption (Fig.~\ref{fig_mass_functions}),
a result of the mass segregation, is very similar to what is observed in the
Arches cluster \citep[][Fig. 14]{Stolteetal02}. This indicates that the
shallowness of the current mass function of the Arches can be explained by
dynamical evolution, without the need for a shallow IMF.

The young compact cluster IRS~13E observed near the GC region provides
strong support for the cluster inspiral scenario. This cluster has a projected
diameter of $\sim 0.5''$ and is at a projected distance of $3.6''$
from Sgr A$^\star$. Its members have a common proper motion with 
velocity $~280\,{\rm km\,s}^{-1}$ 
and no apparent common radial velocity, putting
the cluster on an eccentric orbit. From these data \cite{MPSR04} 
conclude that there is an IMBH at the center of this cluster with
mass $\sim$750--1300\,$M_\odot$. It is possible that these stars are
the central part of a larger cluster as in the five bright stars in
the Quintuplet cluster, but $N$-body simulations show that such a cluster,
consisting of about 2000 stars, will 
rapidly  experience core collapse
($t_{\rm cc} \sim 10^4\,{\rm yr}$; 
H.\ Baumgardt, private communication). Even if there is no IMBH in
the IRS~13E cluster, its presence is a strong indication 
that clustered star formation
took place in the last few million years in the GC region, 
close enough to Sgr A$^\star$ that part of the cluster is now within
the central parsec of the Galaxy.

\cite{AL04} proposed that the young population observed very close
($<0.05\,{\rm pc}$) to Sgr A$^\star$ are stars which reached this region
on radial orbits and displaced stellar mass black holes that reside
there. \cite{GQ03} suggested that a member of this population, S0-2, is the
remnant of a massive binary on an eccentric orbit which got disrupted by
the supermassive black hole. These scenarios naturally complement the
one we investigate in this paper. The stars scattered by the IMBH
will end up on radial orbits and can further interact with and displace
the stellar-mass black holes, or, if they are binaries, they can get disrupted while
on these orbits. This will leave them on less eccentric orbits with smaller
semimajor axes. The orbits of stars in the vicinity of an inspiraling IMBH
have recently been investigated numerically by \cite{LWT05}, who showed that
eccentric orbits with very small apocenter distances are possible if the
inspiraling IMBH is on an eccentric orbit.

Despite these numerical investigations by \cite{LWT05}, the final fate of
a cluster with an IMBH, which disrupts within the central parsec, remains
highly uncertain. More theoretical work is needed to understand both the
final phases of the disruption and the interaction of the IMBH with the
surrounding stars. Further observations of the GC, in particular proper
motion measurements of the closest stars to understand their eccentricity
distribution, and higher resolution observations of the IRS~13E cluster
to resolve its structure and dynamics, will put constraints on the various
scenarios proposed.

\acknowledgements
We thank Tal Alexander and Andrea Ghez for discussions about the mass
distribution near the GC, Holger Baumgardt for providing
his simulation results prior to publication, Marc Freitag, 
Milo\v s Milosavljevi{\' c}, Brad Hansen and Casey Law for 
useful discussions, Emrah Kalemci for help with the
preparation of some of the figures, 
and Ruadhan O'Flanagan for comments on the manuscript.
This work was supported by NASA Grants NAG5-13236, NNG04G176G and 
NSF Grant AST-0206276, and was finalized while MAG was
at the Kavli Institute for Theoretical Physics at the
University of California, Santa Barbara, as a Graduate Fellow.

\appendix
\section{Derivation of Inspiral Rate Formulae}
\label{app_1}
In this appendix we derive the formulae for the inspiral rate $dR/dt$
of a point mass, given a broken power-law Galactic mass distribution
$M(R)\propto R^\alpha$, and an additional central point mass $M_{\rm
SBH}$. We use $R_b=0.38\,{\rm pc}$ to denote the breaking radius,
and the subscripts $1$ and $2$ for regions $R\leqslant R_b$ and $R>R_b$
respectively. We follow and generalize the method of 
\citet[see also \citealt{BT87} \textsection 7.1]{MPZ03}.

For $R\leqslant R_b$, the enclosed mass for a circular orbit is given by
\begin{equation}
\label{eq_mass_dist}
M(R) =  
M_{\rm SBH} + A_1 R^{\alpha_1}
\,.
\end{equation}
The potential corresponding to this mass distribution is
\begin{equation}
\label{eq_pot}
\phi(R) = -\frac{G\,M_{\rm SBH}}{R} + \frac{G\, A_1 R^{\alpha_1-1}}{\alpha_1-1}
\,,
\end{equation}
and the circular velocity is given by
\begin{equation}
v_c(R)^2 = \frac{G\,M_{\rm SBH}}{R} + G\,A_1 R^{\alpha_1-1}
\,.
\end{equation}
Combining these, we obtain the energy per unit mass for a circular
orbit,
\begin{equation}
\label{eq_encirc}
E_c(R) = -\frac{G\,M_{\rm SBH}}{2R} 
	+ \frac{G\,A_1 R^{\alpha_1-1}}{2} \frac{\alpha_1+1}{\alpha_1-1}
\,.
\end{equation}
The acceleration resulting from the dynamical friction on an object of mass $m$ 
is given by \citep[][\textsection 7.1]{BT87}
\begin{equation}
\label{eq_fric_acc}
{\textbf{\em a}}_f = -4\pi \ln\Lambda\, \chi\, G^2 \rho m 
\frac{{\textbf{\em v}}_c}{v_c^3} 
\end{equation}
(see \citealt{MPZ03}, for a description and numerical values of the constants). 
Hence the work done by dynamical friction per unit mass is
\begin{equation}
W_f = -4\pi \ln\Lambda\, \chi\, G^2 \frac{\rho m}{v_c} .
\end{equation}
The time derivative of energy per unit mass in Equation (\ref{eq_encirc}) is
\begin{equation}
\frac{d E_c}{dt} =
 \left(\frac{GM_{\rm SBH}}{2R^2} 
		+ \frac{\alpha_1+1}{2} G A_1 R^{\alpha_1-2}\right)
 \frac{dR}{dt}
\,.
\end{equation} 
By setting this equal to work done by dynamical friction, we obtain
the following expression for orbital decay
\begin{equation}
\label{eq_decay}
\frac{dR}{dt} = 
\frac{-\chi\ln\Lambda\, G^{1/2} A_1 \alpha_1 m R^{(2\alpha_1-1)/2}}
	{(M_{\rm SBH}+A_1 R^{\alpha_1}(\alpha_1+1)/2)
		(M_{\rm SBH}+A_1 R^{\alpha_1})^{1/2}}
\,,
\end{equation}
or in dimensionless form
\begin{subequations}
\label{eq_dim_dec}
\begin{equation}
\frac{d\xi}{d\tau} = 
	-\frac{2\pi\,\alpha_1\,\chi\,\ln\Lambda\, m\, M_1}
		{(M_{\rm SBH}+M_1)^{1/2}} 
	\times\frac{\xi^{(2\alpha_1-1)/2}}{(M_{\rm SBH}+M_1\xi^{\alpha_1}(\alpha_1+1)/2)
			(M_{\rm SBH}+M_1\xi^{\alpha_1})^{1/2}
			}
\,,
\end{equation}
where we have used
\begin{equation}
\xi\equiv\frac{R}{R_0},\;
\tau\equiv\frac{t}{T_0},\;
T_0=2\pi\left[\frac{R_0^3}{G\,(M_{\rm SBH}+M_1)}\right]^{1/2},\;
M_1 = A_1 R_0^{\alpha_1}\,
\end{equation}
\end{subequations}
with $R_0$ being the initial distance from the GC. For $R>R_b$
the equation for the enclosed mass takes the form
\begin{equation}
M(R) = M_{\rm SBH}+A_1 R_b^{\alpha_1}-A_2 R_b^{\alpha_2}+ A_2 R^{\alpha_2}\,.
\end{equation}
Hence, the above formulae can be used for this regime 
by simply substituting $2$ for $1$
in subscripts and replacing $M_{\rm SBH}$ by 
$(M_{\rm SBH}+A_1 R_b^{\alpha_1}-A_2 R_b^{\alpha_2})$. 
However, the above derivations are
valid for $\alpha\neq1$. For $\alpha_2=1$, which is the value indicated by
observations \citep{Genzeletal03}, a slightly
different expression is obtained
\begin{subequations}
\label{eq_spi_in2}
\begin{equation}
\frac{d\xi}{d\tau} =
-\frac{4\pi\,\chi\,\ln\Lambda\,M_2 m}{(M_{\rm SBH}+A_1 R_b^{\alpha_1}-A_2R_b+M_2)^{1/2}} \times
\frac{\xi^{1/2}}
{(M_{\rm SBH}+A_1 R_b^{\alpha_1}-A_2R_b+M_2\xi)^{3/2}}
\,,
\end{equation}
where
\begin{equation}
\xi\equiv\frac{R}{R_0},\;
\tau\equiv\frac{t}{T_0},\;
T_0=2\pi\left[\frac{R_0^3}{G\,(M_{\rm SBH}+A_1 R_b^{\alpha_1}-A_2R_b+M_2)}\right]^{1/2},\;
M_2 = A_2 R_0
\,.
\end{equation}
\end{subequations}
For $R$ given in pc, $A_1=2.46\times10^6\,M_\sun$, and 
$A_2=2.18\times10^6\,M_\sun$ according to values quoted by 
\citet{Genzeletal03}.

\bibliographystyle{apj}

\end{document}